\documentstyle[12pt,epsf]{article}
\begin{document}
\title {Status of the Experiment on the Laboratory
Search for the Electron Antineutrino Magnetic Moment at the
Level $\mu_{\nu} \leq 3 \times 10^{-12}\mu_B$}

\author{B.S. Neganov$^1$, V.N.Trofimov$^1$, A.A.Yukhimchuk$^2$,\\
L.N. Bogdanova$^3$, A.G. Beda$^3$, A.S. Starostin$^3$}
\date{}
\maketitle
{\small
$^1$ Joint Institute for Nuclear Research,141980, Dubna.

$^2$ Russian Federal Nuclear Center - All-Russian Scientific Research
Institute of Experimental\\
$\hphantom{1234}$ Physics, 607190, Sarov.

$^3$ State Scientific Center Institute for Theoretical and
Experimental Physics,  117218, Moscow.}

\bigskip\bigskip\bigskip
\begin{abstract}

The experiment on the direct detection of antineutrino-electron scattering with
an artificial tritium source allows to lower the present-day laboratory limit
for the neutrino magnetic moment by two orders of magnitude. The experiment
brings together novel unique technologies in studies of rare processes of
neutrino-electron scattering:
\begin{itemize}
\item an artificial source of antineutrinos from tritium decay of 40 MCi
activity with the antineutrino flux density $\simeq 6\cdot
10^{14}$ cm$^{-2}\cdot$s$^{-1}$;
\item new types of detectors capable of detecting electrons with energy down to
$\sim$10 eV, namely, a silicon cryogenic detector based on the
ionization-into-heat conversion effect, a high purity germanium detector with
the internal amplification of a signal in the electric field.
\end{itemize}
A compact installation located in a specially equipped laboratory underground
($\leq$100 m w.e.) will provide favorable background conditions for running the
experiment. With the background level about 0.1 events/kg$\cdot$keV$\cdot$day
and detector assembly masses 3kg and 5kg for the silicon and germanium ones,
respectively, the limit for the electron antineutrino magnetic moment
$\mu_{\nu} \leq 3 \cdot 10^{-12}\mu_B$ will be obtained during (1$\div$2) years
of data acquisition. Status of the experiment and state-of-the-art are
presented.
\end{abstract}

\newpage
\section{Motivation}

The possible existence of a neutrino magnetic moment $\mu_{\nu}$ considerably
exceeding the value allowed by the Minimal Extended Standard Model \cite{BWL}
$\mu_{\nu} \sim m_{\nu} \cdot 10^{-19} \mu_B$, ($\mu_B=e\hbar / 2m_e$ being the
Bohr magneton and $m_{\nu}$(eV) being the neutrino mass) is of the fundamental
importance. The prospects for checking the standard model of electroweak
interactions and the search for phenomena outside the limits of its initial
premises are supported by at least two observations, i.e., the solar neutrino
deficit and the anti-correlation of the measured neutrino flux with solar
activity \cite{RD}. A large magnetic moment hypothesis, $\mu_{\nu} \sim
10^{-11}\mu_B$ \cite{VVO}, is so far the unique possibility to explain the
anti-correlation (if confirmed) in the framework of the standard solar model
\cite{BCL}. A number of extensions of the theory beyond the Minimal Standard
Model are proposed, where the required magnitude of $\mu_{\nu}$ can be achieved
independently of a possible neutrino mass \cite{FY}, \cite{V+}.

The present laboratory limits for the neutrino magnetic moment are
derived from the measurement of $\tilde{\nu}$-e scattering in reactor
experiments with electron antineutrinos and are $\mu_{\nu} \leq(1.9\div2.4)
\cdot10^{-10} \mu_B$ \cite {RDP}. More stringent (but model dependent) limits
are found from stellar physics or cosmology, $\mu_{\nu} \leq (0.01\div0.1)
\cdot10^{-10}\mu_B$ (see, e.g., \cite {Raff} for the review). These bounds are
derived from astrophysical considerations against excess cooling of evolved
stars, cosmological considerations for nucleosynthesis, or from SN1987A. For
the time being uncertainties existing in most astrophysical calculations
preclude treating them as reliable constraints \cite{Raff}.

In view of an ample gap between existing experimental constraint and the ones
deduced from astrophysics, it is relevant to lower the present laboratory limit
for $\mu_{\nu}$ below $10^{-11}\mu_B$. The lowest limit expected in current
\cite {Bugey} and standing by \cite {Beda} reactor experiments is $\mu_{\nu}
\leq (0.3\div0.5)\cdot 10^{-10}\mu_B$. The existing projects with artificial
antineutrino sources plan to reach the same level \cite {sour}. A forthcoming
project  has a goal to set a limit for the electron neutrino magnetic moment at
the level $\mu_{\nu} \leq 0.03 \cdot 10^{-10}\mu_B$. The discovery of a
neutrino magnetic moment at this level would reveal the structure beyond the
standard theory and would be influential in the understanding of scenarios with
magnetic-field-induced  spin precession in the Sun, supernovae, active galactic
nuclei, or the early universe.

\section {Idea of the experiment}
Laboratory measurements of $\mu_{\nu}$ are based on the observation of the
antineutrino-electron scattering process. For $\mu_{\nu}\neq 0$ the
differential cross section over the kinetic energy T of the recoil electron is
given by the sum of the standard electroweak interaction cross section (EW) and
the electromagnetic one (EM). At small recoil energies $T \ll  E_\nu$ ($E_\nu$
is the neutrino energy) these two components behave in different ways: the weak
part is practically constant, while the electromagnetic one grows as 1/T
towards lower energies, being practically independent on $E_\nu$ (Fig.1).
Lowering the threshold for recoil electron detection one might choose the
energy interval where the EM contribution to the cross section is larger than
the EW one. This allows improve the sensitivity of the measurements with
respect to $\mu_{\nu}$.

The experiment proposed in \cite {NTY} exploits new unique technologies for
studies of rare processes of neutrino-electron scattering:
\begin{itemize}
\item new types of semiconductor detectors capable of detecting electrons from
neutrino-\\
electron scattering with recoil energy (10$\div$100) eV, where the
electromagnetic scattering dominates over the weak one (Fig.1.);

\item an artificial tritium source (ATS) with the antineutrino flux density
$\simeq~6 \cdot 10^{14}$cm$^{-2}\cdot$s$^{-1}$, which can be achieved in a
compact (1$\div$1.5 liter volume) detector array located inside a thick
cylinder-shell-shaped source.

\end{itemize}

Working with an artificial source one can choose the optimum ratio between
effect-to-background measurement times. With the background level about 0.1
events/keV$\cdot$kg$\cdot$day the limit for the antineutrino magnetic moment
$\mu_{\nu}\leq 3\cdot 10^{-12}\mu_B$ will be obtained during (1$\div$2) years
of data acquisition.

\section { A 40 MCi tritium source}
Choice of tritium as a preferable source of antineutrinos for the $\mu_{\nu}$
measurement is motivated by many physical reasons, of which only few were
mentioned above (see \cite {NTY},\cite {NTYB}). However, to provide the
required antineutrino flux density a 40 MCi ATS is necessary (4kg of tritium).
Up to day as a result of reduction of nuclear weapons a significant amount of
tritium has been stored. The suggestion to use already available tritium for
the fundamental science and, specifically, for the proposed experiment
\cite {NTY}, was recently approved in Russia. The intense tritium source is
presently being developed in the Russian Federal Nuclear Center VNIIEF
(formerly Arzamas-16).

The source being of extraordinary activity, its absolute safety should be
provided at all stages of its life cycle (ATS saturation with tritium, its
transportation, storage and exploitation during the experiment and further
utilization). Some physical, technical and technological aspects of source
designing and construction, as well as safety problems, were considered in
\cite {NTYB}.

The principal requirements to the ATS are:
\begin{itemize}
\item  tritium must be chemically bound to titanium with largest initial degree
of saturation TiT$_{1.9}$;
\item construction should provide an insertion of the cylinder-shaped detector
array;
\item  construction should provide the vacuum-tightness of the inner source
shell and strength reliability 0.999999 during 6 years of the source
exploitation. Compensation of the pressure resulting from titanium tritide
heating by tritium decay or from an accidental heating should be foreseen;
\item construction should enable extraction of the radiogenic helium during ATS
exploitation;
\item ATS should be equipped with a system for the permanent monitoring the
pressure and temperature and a calorimeter, these being the sensors of the
tritium state in the ATS.
\end{itemize}

Besides, the conditions of the low background experiment put forward
additional requirements to the ATS construction materials, procedure of its
manufacturing and maintaining.

\section { Ultra-low threshold semiconductor detectors}
To use all the exceptional advantages of the tritium source for the measurement
of the neutrino magnetic moment novel detectors capable of detecting recoil
electrons at $\sim$10 eV threshold are developed.

{\bf Detectors using the Neganov-Trofimov-Luke (NTL) effect} \cite{NTL}.

Cryogenic detectors have been intensively developed by many groups
recently and have reached a thermal threshold as low as 500 eV per
(150$\div$250) g of the detector mass. These detectors are used to
detect recoil nuclei in the keV range, produced by WIMPS (weakly
interacting massive particles), that are regarded as the most
probable DM (dark matter) candidates \cite {DM}. Despite being an
outstanding achievement in detector technique, this result is
still insufficient for the proposed tritium experiment. A radical
threshold improvement for cryodetectors can be obtained through an
application of the ionization-into-heat conversion phenomenon (NTL
effect) observed in Si and Ge at ultra-low temperatures \cite
{NTS1}. This method can provide a threshold for recoil electron in
the (10$\div$100) eV range, while keeping the thermal threshold
and, consequently, the calorimeter mass relatively large, say 100
eV and 100 g, respectively. The NTL-effect was successfully used
in Dubna for a calorimetric measurement of light absorption
spectrum in silicon at 1K \cite {NTS2} and later observed in a
large volume silicon spectrometer \cite {Tr}.

{\bf Detectors with amplification.}

Another approach is to develop a High-Purity Germanium (HPGe) detector
operating at 77 K with physical amplification of ionization. Presently
germanium detectors are widely used in low background measurements due to the
high purity of germanium crystals: radioactive impurity does not exceed
10$^{-14}$ g/g. The thresholds, (2$\div$10) keV, being mainly determined by
leakage currents and electronic and microphone noises, are too high for the
experiment on the measurement of the neutrino magnetic moment with ATS.

Using internal proportional amplification of the signal one can attain an
effective decrease of the germanium detector threshold. This principle is
realized now in the silicon avalanche photodiodes (APD), where the gain of
about ($10^2\div10^4$) is implemented by an avalanche multiplication of
electrons in the electric field $E\geq(5\div6)\cdot10^5$ V/cm. Such a value
of $E$ is accomplished by a high concentration of impurities in a narrow
junction. As a result the sensitive volume of an APD is only several mm$^3$.

Avalanche multiplication of electrons or holes in HPGe detector of a 100 cm$^3$
sensitive volume can be achieved by the special configuration of the electric
field due to the large difference of cathode and anode sizes \cite {BS}. Such
an avalanche germanium detector (AGD) is designed similarly to a multi-wire
proportional chamber (MWPC). In contrast to MWPC, the electric field in AGD is
defined not only by the applied voltage and electrode dimensions but by donor
(n-type) or acceptor (p-type) impurities concentration as well. The AGD
threshold is defined by the magnitude of the bulk leakage current and for a
planar microstrip Ge detector of a 100 cm$^3$ volume the threshold
$E_{th} \simeq $10 eV is expected \cite {BS}.

A prototype avalanche germanium strip detector of a 20 cm$^3$ volume is
manufactured now.

\section { Experimental installation}

When a more detailed consideration of the future installation units began, it
was understood that an ATS of the cylinder-shell-layered shape would the most
adequately meet some technological requirements of its manufacturing and
maintaining. Calculations showed that the antineutrino flux density inside the
cylinder of external diameter D=30 cm is of the same order of magnitude as
inside the sphere considered in \cite{NTY}. This allowed consideration of the
future installation design for the cylinder-like source geometry. The final
optimizing of the source dimensions can be done after the detectors effective
size is determined. Due to the low end-point energy of the tritium decay
spectrum ($E_0$=18.6 keV) no special passive shielding between the ATS and
detectors is needed: the bremsstrahlung is absorbed within the source.

The spectrometer including AGD must have a (4$\div$5) kg mass, it can be
fabricated from (5$\div$7) separate modules of a 150 cm$^3$ with a mass of
about $\sim$0.8 kg each. Cryogenic silicon detectors of a 100 cm$^3$ volume,
mounted (14$\div$16) on a stack, provide a net mass of about 3 kg.

The scheme of the installation is shown in Fig.2. For the
shielding the classical scheme is proposed: an air-proof 5 cm
thick container of low-background copper surrounding the source is
followed by a 8 cm thick layer of borated polyethylene and a 15 cm
thick layer of lead. External plastic scintillator of a 4 cm
thickness vetoes charged cosmic components. Gaseous nitrogen
circulating around the copper container removes air-born
radioactivity (Rn). The cryostat cup made of low-background copper
houses the detectors. The low temperature of the dilution
refrigerator (for cryodetectors) (not shown in Fig.2) or of the
nitrogen Dewar (for the AGD) is transferred to the detector by a
cold finger.

Construction of the shield and of the ATS support (not shown in Fig.2) must
allow access to and extraction of the tritium source.

A compact installation will be located in a specially equipped laboratory
underground ($\geq $100 m w.e.) providing favorable background conditions.

Development of two types of detectors for antineutrino-electron scattering
detection opens a perspective of simultaneous running two independent
experiments with the same ATS. An identical set up but with deuterium instead
of tritium will be constructed and located in the same site nearby to measure
the background. While one spectrometer measures the effect + background (data
acquisition during 50\% of total experiment duration proves optimum), the other
one measures the background with its deuterium-filled twin and {\it vice
versa}. This would enable control operation of both spectrometers and would
essentially increase the statistics collected during ATS effective functioning.

\section { Background}

The main sources of the background in the future experiment are: environmental
radioactivity, intrinsic contamination of the ATS and shielding materials,
intrinsic contamination of the detector (including the cosmogenic component,
especially for Si cryodetectors), airborne radioactivity (Rn), cosmic
radiation, neutrons from natural fission ($\alpha$, n) reaction. Traditional
and specially developed background suppression methods will be used:

\begin{enumerate}
\item deep underground operation ($\sim$ 100 m w.e.). Muonic flux at this
depth is about 2 $\mu/$m$^2\cdot$s, and background from the secondary cosmic
neutrons is of the same magnitude as that of the environmental ($\alpha,n$)
radioactivity;
\item passive shielding reducing external radiation and neutron backgrounds;
\item  material selection - using radiation pure materials for detectors, ATS,
passive and active shielding reduces the background of the installation;
\item active background discrimination (veto and coincidence techniques). Veto
with using the plastic scintillator discriminates charged particles passing
through the detector and neutrons correlated with muon capture in the
installation. Anti-coincidences between the separate modules of the detector
array suppress the radiation background;
\item pulse-shape analysis methods to suppress microphone and electronic
noises.
\end{enumerate}

Monte Carlo simulations can be used for understanding the structure of the
background with a goal of its reduction. The quality of computer intense
modeling depends on a detailed knowledge of the experiment geometry, location
of the sources of characteristic radiation, and complete consideration of all
physical processes involved. In the experiment on the neutrino-electron
scattering the background for the single-electron events, i.e., recoil
electrons with E $\leq$ 1000 eV, comes mainly from the following processes:
photons with energy E $\leq$ 1000 eV, Compton electrons E$_e \leq$ 1000 eV,
electromagnetic scattering of neutrons on electrons, nuclear recoils from
neutron-nucleus scattering.

The radiation background in semiconductor detectors, which plays a decisive
role in the future experiment, has been well studied for the region above 2 keV
in dark matter searches \cite{CDMS}. The lowest measured background is 0.08
events/keV$\cdot$kg$\cdot$day for Ge detectors \cite {DM}. For Si detectors
the radiation background is somewhat larger.

Since technologies available now allow obtaining tritium of extremely high
degree of purification, the main attention should be paid to titanium chosen
as  a tritium carrier in the ATS. Monte Carlo calculations of the background
due to radioactive contamination of Ti by the U-Th chain and $^{40}$K were
performed. This component should not exceed the background $\sim$ 0.1
events/keV$\cdot$kg$\cdot$day. Then the allowed level of Ti radioactive
contamination proved to be $\leq 10^{-10}$ g/g. Industrial titanium is of the
purity $\sim(10^{-9}\div10^{-8})$ g/g. So special efforts are needed to provide
its necessary radioactive purity.

Concerning the correlated background, note that tritium antineutrino coherent
scattering on nuclei is inessential, producing nuclear recoils of fractions of
eV.

\section { Expected results}

Expected rate of antineutrino-electron magnetic scattering events per day
for two values of $\mu_{\nu}$ (effect) and of weak scattering events are shown
in the Table for two energy intervals of recoil electron detection.

\hfill  {\bf Table}
\begin{center}

{\bf Number of $\tilde{\nu}$-e magnetic ($N_M$) and weak ($N_W$) scattering
events and background (B.g.) expected per day for different energy intervals
of detected recoil electrons}

\vspace*{0.5cm}

\begin{tabular}{|l|l|l|} \hline

Energy interval (eV)         &  10$\div$200  & 10$\div$1260 \\
 $N_M$($\mu_{\nu} = 1 \cdot 10^{-11} \mu_B$) & 1.4 & 2.4\\
 $N_M$($\mu_{\nu} = 3 \cdot 10^{-12} \mu_B$) & 0.13 & 0.22\\
 $N_W$&  0.04 &  0.15 \\
B.g. &   0.1 &  0.5 \\ \hline
\end{tabular}
\end{center}

The antineutrino flux density is taken $\simeq 6 \cdot 10^{14}$ cm$^{-2}\cdot$
s$^{-1}$, the detector mass is 4 kg. In calculations the electron binding in
the atom was taken into account \cite {MIK}. For the more detailed
consideration effects of atom binding in the crystal should be included. The
numbers of background events were calculated assuming the background level to
be 0.1 events/keV$\cdot$kg$\cdot$day.

It is clearly seen from the Table that at low threshold the number of expected
events changes inessentially when the effect is observed for (10$\div$200) eV
region compared to the total recoil energy range (10$\div$1260) eV. At the same
time narrowing energy range of recoil electron detection to (10$\div$200) eV
reduces both non-correlated and correlated background (weak interaction
contribution) noticeably. Of course, the assumption about the uniform
background below 1 keV should be carefully checked, which is the primary task
in the measurements.

The sensitivity of the experiment for the neutrino magnetic moment can be
determined using data from the Table. Assuming the total duration of the
measurements to be 400 days (200 days with ATS and 200 days of background
measurements) the achievable limits are $\mu_{\nu}\leq 2.5 \cdot 10^{-12}\mu_B$
for the energy interval 10-1260 eV and $\mu_{\nu}\leq 2.2 \cdot 10^{-12}\mu_B$
for (10$\div$200) eV at 95\% C.L.

\section {Status and conclusions}

The Program "Measurement of the neutrino magnetic moment at the level
$\mu_{\nu}\leq (1\div3)\cdot 10^{-12}\mu_B$" has been approved by the Ministry
of Atomic Energy of Russian Federation (Minatom). R\&D on the ATS, two types of
detectors and all relevant problems have begun in JINR, ITEP and RFNC VNIIEF.

Experimental study of neutrino properties and interactions with matter is a
challenge for low energy physics. Discovery of a neutrino magnetic moment at
the level $\mu_{\nu}\leq 3 \cdot 10^{-12}\mu_B$ would indicate physics beyond
the standard model of electroweak interactions and would radically change the
modern astrophysical scenario. In particular, these results would impact our
understanding of the observed variations of the solar neutrino flux. The novel
detector technologies developed for this experiment can further be considered
for other elementary particle physics research (dark matter search, neutrino
coherent scattering on nuclei, solar neutrino measurements) and for other
fields of fundamental and applied physics requiring detection of low energy
particles.

\section {Acknowledgment}

This work was performed under the Contract 66.04.19.19.00.802 with the
Department of Atomic Science and Technology of Minatom.

Authors appreciate useful discussions with Professor V.G.Zinov and especially
his suggestion on the ATS shape. Authors are grateful to E.V.Demidova for help
in calculations.

{\bf Figure captions.}

Fig.1. Differential cross sections of the $\tilde{\nu}-e$ scattering over
electron recoil energy for $^3$H emitter. Contribution from magnetic scattering
is shown for $\mu_{\nu}=m\cdot 10^{-12}\mu_B$ ($m$=1,3,10). Dashed line shows
the standard electroweak cross section. Arrow indicates energy threshold for
existent semiconductor detectors (SCD).

Fig.2. Principal layout of the installation for the $\mu_{\nu}$
measurement with the ATS.

\end{document}